**Development and Evaluation of Dental Image Exchange and Management System: A User-Centered Perspective**

**Short title:** User-Centered development of dental exchange radiology system

Rahimi B[1], Karimian S[2 *], Ghaznavi A[3], Jafari Heydarlou M[4].

1) Professor of Medical Informatics Department of Health Information Technology, School of Allied Medical Sciences Urmia University of Medical Sciences, Urmia, Iran.

bahlol.rahimi@gmail.com

2) Student Research Committee, Urmia University of Medical Sciences, Urmia Iran

sajjad.karimian94@gmail.com

*corresponding author

3) Assistant Professor, Department of Oral & Maxillofacial Radiology, Dental Faculty, Urmia University of Medical Sciences, Urmia, Iran.

aisanghaznavi@yahoo.com

4) Assistant Professor, Department of Oral Medicine, Dental Faculty, Urmia University of Medical Sciences, Urmia, Iran.

dr.jafarymo@yahoo.com

**Abstract**:

***Introduction***: Systems that exist in the hospital or clinic settings are capable of providing services in the physical environment. These systems (e.g., Picture Archiving and communication systems) provide remote service for patients. To design such systems, we need some unique methods such as software development life cycle and different methods such as prototyping. Clinical setting: This study designs an image exchange system in the private dental sector of Urmia city using user-centered methods and prototyping. ***Methods***: Information was collected based on each stage's software development life cycle. Interviews and observations were used to gather user-needs data, such as object-oriented programming for developing a Prototype. ***Results***: The users' needs were determined to consider at the beginning. Ease of use, security, and mobile apps were their most essential needs. Then, the prototype was designed and evaluated in the focus group session. These steps continued until users were satisfied in the focus group. Eventually, after the users' consent, the prototype became the final system. ***Discussion***: Instant access to Information, volunteering, user interface design, and usefulness were the most critical variables users considered. The advantage of this system also includes less radiation to the patient due to not losing and missing the clips of the patient's images. ***Conclusion***: The success of such a system requires the consideration of end-users needs and their application to the system. In addition to this system, having an electronic health record can improve the treatment process and improve the work of the medical staff.

**Introduction**:

Considering the systems in health-related contexts, such as Hospital Information systems (HIS), Picture Archiving and Communication Systems (PACS), and Radiology Information systems (RIS), they are mainly related to patients and health providers at a physical site, for instance, a hospital or clinics. Such systems are primarily designed to provide a comprehensive service at a physical site (1,2). In this regard, systems such as RIS and PACS have the specific potential to transmit patient textual and visual information in the healthcare setting (3,4). Despite all benefits of web-based image transition systems, there are some problems in developing these systems, including privacy, security, cost-related issues, usability, evaluation, infrastructure, and accessibility in terms of hardware. Finally, all users have network connections through mobile phones (3,5–7).

Human-Computer Interaction (HCI) is one of the effective methods used for successful information systems analysis. Focusing on these kinds of analysis is mainly related to user-effected factors, including behavioral, cognitive, easy to use, and ergonomic ones. One of these approaches is the 'User-Centric Approach' in software analysis (5–8) which primarily considers the affairs from the users' perspective. The system's functions regarding the type of users using this system based on their different approaches are the most critical items that should be addressed in system analysis. This information should be widely gathered from all stakeholders and system end-users to create a system that meets the complex needs and expectations of the organization and its users regarding existing processes (5–9).

Furthermore, such systems must be evaluated for validation, properly working, cost analysis, or any other goal that the system sets to achieve (10). Without consideration for the needs of all users in designing and implementing any system, the system would not work well at all. It will be tackled with colossal failure (4,7), which is highly important in evaluation (5,6,10).

In general, a lack of paying attention to users' needs concludes the failure of the Electronic Health Record (EHR) system (11). Effectively, one of the main reasons for the success of a low-cost portable system is the involvement of users in the early stages of prototype design (12). Some studies reveal that focusing on users and their needs plays an essential role in successfully implementing all systems (13–15). Due to the increasing use of exchanging medical information systems, the importance of users' comments in designing, implementing, and accepting the type of systems is highly significant (16). To design such effective systems, system developers, analyzers, and designers should use some methods, including waterfall methods, prototype development, and other methods, to develop information systems appropriately and evaluate the working system (5,6,10).

Appropriate design and creation refer to the Software Development Life Cycle (SDLC) in software or system design. These methods are used to design a system or software and start conducting several stages of designing and planning, programming, implementing, and maintaining the required system or software by guiding the management and executive team. The success and advantage of such systems are considered by evaluating the needs analysis of users, mainly end-users, designed by system designers and programmers (15).

Due to the central fact that such a system does not exist in Urmia dental private sector clinics, this study aims to design, develop and evaluate a user-centered system. Also, the importance of optimal use of available resources for a fair exchange, evaluate its impacts on stockholders efficiency (e.g., physicians, patients, and assistants), and data management of oral and maxillofacial radiology images in Urmia dentistry private sector clinics.

**Materials and Methods:**

The present system is designed and developed in a repetitive methodology based on prototype development. In the first phase, the 'users' requirements' are gathered regarding ethnographic methods such as conducting interviews and observation to design and develop the required system. The main reason for conducting an interview is to gather information regarding the system's features, user interface, user needs, and users' expectations of the system. Regarding the observation, it is mainly used to know users' workflow and their activities. In phase two (system design), the creation of a prototype with Object-Oriented Programming (OOP), Unified Modeling Language (UML) analysis, and prototype software model based on the gathered Information from phase one is highly considered. The result of phase two leads to the production of a system prototype. Phase three (prototype implementation) will evaluate the prototype in a focus group session and reiterate the last two processes to make the final prototype as the system. Lastly, in phase four (system implementation and evaluation), the system will be implemented in the real world with real users and eventually evaluated for usability.

*Phase one:*

In conducting the study, User-centered approaches such as interviews and observation are used to get user requirements. A face-to-face semi-structured interview was used to fully understand users' needs and what they expect from the system. To do so, 25 academic staff at Urmia University of Medical Sciences specialized in Endodontics, Pedodontics, Periodontics, Oral Medicine, Orthodontics, Restorative & Esthetic Dentistry, Prosthodontics, Oral & Maxillofacial Surgery, and Oral and Maxillofacial Radiology attended at this study. From each specialized group, two persons were selected. Five dentists, radiology assistants, and patients participated in the interview. Interviewees were selected based on their experience and eagerness to participate in this study. Each interview took 20-35 minutes, and four aspects of the used system, including features of the system, User Interface, public oral health education, and users' workflow, were considered (Table 2). These questions were considered an essential aspect of the design and were designed by the research team. Before conducting any interview, getting permission from all participants to record their voices and transform them into textual data was highly considered in this study. The interview was conducted by one of the researchers who had previous knowledge of software development. Finally, the observation method was used to gather Information regarding dentists' workflow in their private clinics and how they order exams.

*Phase two:*

The Prototype methodology is used to design the required prototype according to the users' needs in the previous section. Accordingly, after extracting the Information through interviews and observation, a required prototype is available to get feedback and comments more quickly. UML analyzes users' needs, and the Use Case Diagram is drawn to visualize their needs and expectation. Finally, by conducting Object-oriented programming, a prototype is developed. Afterward, the Entity-Relationship Diagram (ERD) is drawn. In this phase, the prototype design is conducted regarding the obtained Information from the users' expectations and needs.

This web-based prototype is mainly designed by applying Hypertext Preprocessor (PHP) as an open-source dynamic scripting language. PHP develops various frameworks for programming cake PHP, lithium, and other frameworks. This framework helps programmers with better data management, security, and coding. In this scenario, a self-made framework for developing this prototype is used (17). The architecture of this work followed Model-View-Controller (MVC) for better controlling databases, viewing the Information, and practical coding (18). the Bootstrap framework developed and presented by Twitter is used To display this web application on various devices. This framework allows the programmer to display this web system on different screen sizes, such as mobile phones, tablets, laptops, and other devices (19). Due to the urgent need for a faster system, Asynchronous JavaScript and XML, or so-called AJAX technology, are highly recommended and applied to speed up the display. Without having to refresh, this technology can display Information and events on the client-side (20). For security affairs, JSON Web Token (JWT) is highly used in API for a secure connection in the retrofit system. Also, SSL (Secure Socket Layer) connection is applied for secure data transmission between API and other devices.

Because of a necessary need for a mobile application to connect this system to mobile applications, the Application Program Interface (API) is applied. This link will create a web application for the mobile platform (21). In this study, the targeted mobile operating system was the Android open-source operating system, not IOS, because of the limited budget. This web-based system is designed to run on Android mobile devices by installing a program to display Information to the dentist and carry out a limited number of features. The database used to store data was the MySQL database which is free of charge.

*Phase three:*

Accordingly, after designing the prototype, its implementation regarding the previous phases is performed. Active users during interviews are selected and gathered in a focus group meeting. The prototype system is presented in this session. Users can interact with this prototype, view the system user interface, and mention their comments in this regard. After this phase is done, their comments are added. Afterward, users agreed with this prototype as their final system. After designing, this web system runs on a local Linux web server displayed in the focus group session. After re-designing and re-providing users' requirements, the problems or computer bugs are critically examined by the programmer and analyst of the system.

*Phase four:*

As the final stage of any system's life cycle, the final system will be implemented in the real world with actual patient data. At first, creating a user account for dentists and radiologists is needed. Then, radiologists and dentists were individually trained in a separate session before using the system. Staff was given the training needed to use at a radiology center. On the first day, four dentists entered the system. Then, in the following days, the number of dentists visiting the radiology center increased to eight. Finally, we have introduced a second radiology center to the system on the first weekend (dentists are no longer in the office on Thursdays, and there will be lower radiologists' workloads). Like the first center, a small number of dentists' patients entered the system. Finally, four radiology centers entered within four weeks to familiarize themselves with the system's workflow and data entry.

The process was admitting the patients associated with the dentists who are members of the system. The new process was explained to patients and given an informed consent form. After taking the radiologic image, their radiography images are sent through this system to their dentist. Patients' hard-copy images are given to themselves with usernames and passwords to access their soft-copy images via the system provided. On the other hand, dentists have access to their patient's information through the system. If they need any help from other dentists or radiologists, they can refer the patient.

During the first month of implementing this system, an evaluator did a formative evaluation to prevent the system from any system failure. Finally, after the implementation of this system, a self-made questionnaire based on the Technology Acceptance Model (TAM), Unified Theory of Acceptance and Use of Technology (UTAUT), Task-Technology Fit (TTF), and diffusion of innovations theory to evaluate the system was designed (22,23,32,24–31). The questionnaire contains three parts, demographic Information, IT knowledge, and system evaluation that include efficiency, ease of use, quality of UI, quality of communication, reliability, system satisfaction, and voluntary usage of the system. Data were analyzed with SPSS 16 and tested with backward multivariate linear regression. The validity of this questionnaire was checked by four faculty members of the Department of Health Information Technology (HIT) and two faculty members of dentistry. Also, the reliability of the questionnaire was cheeked by test-retest (Cronbach's alpha).

**Results:**

*Phase one:*

Interviews results: Demographic Information of interviewees is demonstrated in Table 1. The lists of questions for the required interviews were categorized in Table 2. the collected data from interviews were categorized into four main aspects. Afterward, the results were ranked based on the repetition rate (low, medium, high, very high), also shown in Table 3.

Observation results: According to the observations, the workflow in the clinics is to A) visit the patients, B) order radiography if necessary, and finally, C) bring back the radiography image/images to his/her dentist (in Urmia, the radiology dental sector is separated from other sectors in oral health care wards). In some clinics, imaging radiography is available only for periodical images but not for other types of dental imaging if and only if the principles of radiation protection are required.

Put Tables 1, 2, and 3 here.

*Phase two:*

This section presents the logical and physical design of the prototype and system. The use-case diagram explains what users of the system can do and the actions they can make.

Login pages for different users have a different dashboard that briefly presents their work status, such as unread messages number of patients. Radiologists' dashboard has features such as adding patient's name, last name, national code, address, doctor, birth date, phone number, disease history, and insurance code for data entry. However, only their name and national code are mandatory. After adding a patient, it will be added to the patients' list, and by clicking on their name, their radiography image and its report can be added, respectively. Also, on that list, any referral patient can be seen. In the message section, they can send a message to the system administrator as a help desk. This dashboard is demonstrated in figure 1.

Put figure 1 here.

Dentists' dashboard (demonstrated in figure 2) has features including their patients and their images. Also, they can send their patient's images to any colleague or can see their colleagues' referral patient's radiography and its report. Also, all pages can be viewed on mobile browsers without any complications, as shown in figure 3.

Put Figure 2 here

Put Figure 3 here

*Phase three:*

Selections of significantly active users in the requirement phase are made as to the focus group (10 people). First, their needs and expectations are considered, then the prototype is presented respectively. They have interacted with this system, see its user interface, and mentioned their opinions. Due to simple access to mobile, the demand for a developed mobile app (n>6) is highly recommended. Furthermore, the need for radiological reports and their graph is mainly requested (n>6) as a part of the system. Other significant demands have a referral system for radiologists, adding radiologic exams for patients, and creating a mark on the photo to easily convey their meaning to the radiologist or dentist's colleague (3>n>5). Finally, after adding these requests, the system is re-presented to another focus group, and users express their range of satisfaction. Their opinions regarding system operation, data entry, minimum data set, and system workflow are evaluated and considered by the users working on the system.

*Phase four:*

In this study, four radiology centers, nine dentists, and 50 patients entered the system in first month of implementation. Cronbach's alpha is 0.986. The correlation between the variables is show at Table 4, and multivariate linear regression is shown at Table 5 respectively. End-user including 22.5% dentists (n=9), 10% radiologists (n=4), 52.5% patients (n=21), and 15% radiologist's assistants (n=6) willingly completed the questionnaire (total n=40). The participants age is $33.30 \pm 6.42$ (Mean $\pm$ STD). Degree of education are 17.5% diploma (n=7), 37.5% bachelor (n=15), 12.5% master (n=5) and 32.5% Ph.D. (n=13) and field of study are 5% Periodontist (n=2), 22.5% Radiologist (n=9), 5% Orthodontist (n=2), 5% Oral Medicine (n=2), 7.5% General Dentist (n=3) and 55% other (n=22). Also 50% of the users have had training for using the system (n=20), and 50% hadn't any training (n=20). Among trained users 50% had in person training for using the system (n=20). 90% of users use the system less than seven hours in a week (n=36) and only 10% use the system more than 21 hours in a week (n=4) that mostly are radiologists and their assistants. Users' device is 52.5% PC via the website (n=21), 40% mobile phone (n=16) and 7.5% with tablet (n=3).

Put Table 4 here.

Correlations between these variables are significant (CI=95%, P-value <= 0.05):

- Beneficent with UI, Interaction, reliability, satisfaction, voluntary and IT knowledge
- UI with beneficent, interaction, satisfaction, and IT knowledge
- Interaction with beneficent, UI, reliability, satisfaction, and voluntary
- Reliability with beneficent, interaction, satisfaction, and voluntary
- Satisfaction with beneficent, UI, Interaction, reliability. voluntary and IT knowledge

- Voluntary with beneficent, Interaction, reliability, and satisfaction.

The different number in the correlations table is because some parts of the questionnaire for patients and assistants were not asked. As a result, the N is 19.

Put table 5 here

**Discussion**:

In this study, the users' needs regarding their working settings in Urmia City are divided into four general categories of needs. In the first category, the features of this system, including descriptions of users' needs, comments, suggestions, communication among colleagues, images or reports sent to physicians, use of a mobile application, and sending messages to doctors are mainly considered. The second feature in designing the system is appearance and users' interface, which are other important factors. Ultimately, all users have been asked about the existence of such a system in their work. Also, the patients' economic and welfare point of view, which has proclaimed the system's success, is comprehensively considered.

According to the interview results obtained from dentistry faculty members at Urmia University of Medical Sciences, the effect of a mobile application accompanied by a report of images, the convenience of working with the system, and a user-friendly interface regarding users' needs were evaluated. In this regard, speed, security, communication with others, notifications, 2D images, proper design, general education for the patient, patient access, orders sending through this system, image processing, and usefulness in the user's work routine are highly demanded analysis. Meanwhile, the need for paramedical Information, 3D imagery, and an online booking system has been a minor requirement in users' needs analysis. These results indicate that dentists, radiologists, and their assistants are looking for a system that works quickly to reveal the results of patients without needing any prior training. In this regard, easy accessibility to mobile is the main reason for its popularity among users. Using free programs and methods such as Linux Web Server, PHP programming language, Android application, and MySQL database to cut down the project costs are considered in this study.

In a User-Centric method of designing a system, the primary process is to consider each stage of the system development, including analysis, development, and evaluation, respectively. In analyzing the system, using ethnographic methods to assess users' needs by interviewing, studying the field, preparing the workshop for a focus group, and analyzing all activities to get a result should be considered. System design, a variety of development methods such as Rapid Implementation Method, Prototype Approach, using software-based system analysis methods including Use case to analyze and display users' needs is highly recommended. This method requires no prior computer knowledge for users. Finally, after analyzing and developing systems, a system evaluation method is used to evaluate this prototype or system from the users' point of view quantitatively (14,15).

Using interviews to verify users' needs and converting this Information into use-case information for software programming are all applied in this research. The present study considers the dentists' interviews and takes all system stockholders' comments to have an effective system. After developing and injecting user views into the prototype and final system, the system must be tested in a real-world environment with factual Information by the end-users involved in its design for a month.

Pearson's correlation analysis indicates that the ease of using a variable has little effect on the other variables. However, the beneficent, voluntary item and IT knowledge variables affect other variables. This deduction may mean that end-users do not feel that the designed system is helpful for them, voluntarily, not compulsively. Ultimately, IT knowledge can be expressed as accepting the system for them, and it is easier and better to work with the system. According to studies, the success of a system is relevant to engaging its users (33). Also, the regression relationship between the evaluation variables is evident. The most impact on the variables is voluntary, and most of the variables with a coefficient determination of about one. Voluntary use of the system affects the amount of beneficence, ease of use, system interaction, reliability, and satisfaction. The user interface is related to the degree of usefulness and interaction, in which beneficence is associated with the user interface and voluntary, and the system interactions are associated with voluntary and satisfaction. In general, volunteering, satisfaction, and user interface have impacted the system's design most.

Users who commented on the use of this system after using it also stated that they had the proper hardware infrastructure to use such systems in the health sector. Not having a suitable system in the office for quick viewing of images, the lack of uploading images from the radiology center images due to the high workload, aging process, and not being accustomed to viewing digital images in the office are disadvantages of the system from the end user's perspective. Regarding the advantages, eliminating medical clips, storing images in a centralized environment, reducing the probability of clips being lost or destroyed, and less patient exposure are mainly emphasized. This system will improve their workflow, and, in some cases, the lack of electronic, oral health records in dentists' offices was the main problem in this study. If there are some rules to help create a fully electronic oral health record, they would be excellent assistance in removing all physical data in oral health sectors. It also reduces printing and maintenance costs for radiology centers, and thus patients' imaging costs will be reduced.

**Conclusion**:

In general, the success of the IT system in the private dental sector in Urmia City with providing the appropriate infrastructure and designing a user-centric system and voluntary deployment of these systems can provide cost and health benefits to the stockholders of this sector. Also, it can improve the patient's treatment process by having an electronic health record in the private sector to have a cooperative relationship with the governmental sector.

*Limitation:* An Android mobile application was developed due to having a problem publishing the IOS app. Furthermore, there is a lack of patient corporation with the research team.

**Acknowledgment:**

This study was extracted from a Master of Science thesis and funded partially by the Urmia University of Medical Science [grant number IR.UMSU.REC.1397.308]. We extend our warm thanks to the faculty member of the Dentistry School at Urmia University of Medical Sciences, who provided high insight and expertise in conducting the research**.**

**Declaration of interest statement:**

The authors report there are no competing interests to declare

**References***:*

and system success. Inf Softw Technol [Internet]. 2015 Feb;58:148–69. Available from: https://linkinghub.elsevier.com/retrieve/pii/S0950584914001505

Appendix 1:

**Part I**

**General specifications:**

- Level of education:
    - Diploma
    - Bachelor
    - Masters
    - Dentist
    - PHD
- Field of study………….
- Gender
    - Male
    - Female
- Received Training to use the system
    - Yes
    - No
- In case of positive answer, training downloaded from the website or a person trained you?
    - Downloaded from the website
    - A person trained me
- How used the system?
    - Mobile phone
    - Tablet
    - Website
- The average daily hours spent working with the system is …………….
- System usage during the week……………….

**second part**

**Knowledge of information technology**

| Questions | Very much (5) | Much (4) | Medium (3) | Low (2) | Very little (2) |
|---|---|---|---|---|---|
| **In general ,the skills needed to work with the Microsoft Office applications.** | | | | | |
| **The level of your ability to properly install a software on a computer.** | | | | | |
| **Awareness of the benefits of using an Information system in a health care setting** | | | | | |
| **Average usage of computer and mobile phone to facilitate your daily affairs** | | | | | |
| **Daily work usage with computer and mobile at work** | | | | | |
| **Email sent and received in the workplace** | | | | | |

**Third part**

**System Satisfaction Information**

| NO | Questions | Very much (5) | Much (4) | Medium (3) | Low (2) | Very little (2) |
|---|---|---|---|---|---|---|
| | **A) Usefulness** | | | | | |
| 1 | The system saving time in retrieving the images | | | | | |
| 2 | The system improved access to medical resources (clinical images) | | | | | |
| 3 | The system fulfilled my clinical needs | | | | | |
| 4 | The system is beneficial in my work | | | | | |
| 5 | The system improved my work performance | | | | | |
| 6 | The system improved my work efficiency | | | | | |
| 7 | The system generally improved my work performance | | | | | |
| 8 | The system learned how to use both usage of technology and clinical information | | | | | |
| 9 | The system information is up to date | | | | | |
| | **B) Ease of use** | | | | | |
| 1 | The system is easy to use. | | | | | |
| 2 | Learning how to use the system is easy. | | | | | |
| 3 | I get my needed information quickly | | | | | |
| 4 | Using the system does not require much mental effort | | | | | |
| 5 | Interactivity with the system is clear and understandable | | | | | |
| 6 | The system's response time is fast | | | | | |
| 7 | Using a digital system is better than a paper system. | | | | | |
| 8 | Data input is easy | | | | | |
| | **C) UI quality** | | | | | |
| 1 | The system is attractive | | | | | |
| 2 | I like to use the system | | | | | |
| 3 | The system is simple and understandable | | | | | |
| 4 | I like the system color scheme | | | | | |
| 5 | The system display in mobile is good | | | | | |
| 6 | In the system, error and success messages are well displayed. | | | | | |

| NO | Questions | Very much (5) | Much (4) | Medium (3) | Low (2) | Very little (2) |
|---|---|---|---|---|---|---|
| 7 | The output format of the information in the system is appropriate | | | | | |
| 8 | Appropriate colors have been used for different parts of the system. | | | | | |
| 9 | There is simplicity in designing system pages | | | | | |
| 10 | There is a logical connection between the pages of the system. | | | | | |
| | **D) System communication quality** | | | | | |
| 1 | I can communicate with other colleagues. | | | | | |
| 2 | Communication with other colleagues is easy to find. | | | | | |
| 3 | It is possible to communicate with site manager | | | | | |
| 4 | Communication with site manager is easy to find. | | | | | |
| 5 | The system search function works well. | | | | | |
| 6 | I can easily get help when I have a problem with the system. | | | | | |
| | **E) Reliability** | | | | | |
| 1 | The images in the system are as good as the physical images. | | | | | |
| 2 | The images in the system are as accurate as the physical images. | | | | | |
| 3 | If I get an error or problem, I can easily fix it. | | | | | |
| 4 | The system provides me with a proper message to fix the error. | | | | | |
| 5 | The error message provided by the system is useful | | | | | |
| 6 | The system help is complete | | | | | |
| 7 | System security is reliable | | | | | |
| 8 | The information content of the system meets my needs. | | | | | |
| 9 | The information provided by the system is sufficient. | | | | | |
| | **F) System satisfaction** | | | | | |
| 1 | I feel comfortable communicating with colleagues through the system. | | | | | |
| 2 | The system is an acceptable way to receive clinical images. | | | | | |
| 3 | Using the system is exactly the same as before | | | | | |
| 4 | I have a positive feeling about using the system. | | | | | |
| 5 | The quality of work in providing services has been improved by using the system. | | | | | |
| 6 | Using the system is essential in my job | | | | | |
| 7 | Using the system is related to my job | | | | | |
| 8 | Using the system does not increase my workload in the workplace | | | | | |
| 9 | In general, I am satisfied with the system. | | | | | |
| | **G) Volunteering** | | | | | |
| 1 | I use this system voluntarily. | | | | | |
| 2 | Using this system, by my colleague's opinion is required | | | | | |

| NO | Questions | Very much (5) | Much (4) | Medium (3) | Low (2) | Very little (2) |
|---|---|---|---|---|---|---|
| 3 | Using this system in my department is voluntary. | | | | | |
| 4 | If the system presented in university, I will use it. | | | | | |
| 5 | I will also use the system to care for patients in the normal course of work. | | | | | |
| 6 | I will use the system in clinical and non-clinical usages | | | | | |
| 7 | I will use the system to care for my patients continuously. | | | | | |
| 8 | I will use such a system again. | | | | | |
| 9 | I recommend using this system to others. | | | | | |
| 10 | I think others should use this system as well | | | | | |

Tables:

| Group | Number | Age (AVG+STD) |
|---|---|---|
| Dentists | 20 | 38.69 ± 5.95 |
| Assistants | 10 | 31.43 ± 9.14 |
| Radiology assistants | 5 | 35.80 ± 4.60 |
| Patients | 5 | 24.20 ± 0.84 |
| Total | 40 | |

**Table 1.** Demographic Information of end-users

| Aspects | Questions |
|---|---|
| The features of this system. | a. What do you expect from the system in general?<br>b. Would you like to have collaborated with other dentists and your patients?<br>c. Would you like to transmit only the picture or anything else?<br>d. Would you like to run the system on smartphones?<br>e. Would you like the system to send you notifications? |
| User interfaces. | a. What would you like the system user interface to look like?<br>b. Would you like a different color scheme for the system? |
| Public oral health education | a. Would you like the system to educate the patients in a multimedia context? |
| Dentists and radiology workflow | a. Would you like the system to help you with your tasks? |

**Table 2.** four aspects of interviews questions asked from users

| Aspect | Sub-feature | Low (n<10) | Medium (11<n<19) | High (20<n<30) | Very high (n>31) |
|---|---|---|---|---|---|
| System Feature | 1-Speed | | | ✓ | |
| | 2-Security | | | ✓ | |
| | 3-Communication | | | ✓ | |
| | 4-Notification | | | ✓ | |

| | | | | | | | | |
|---|---|---|---|---|---|---|---|---|
| User Interface | 5-Image processing | ✓ | | | | | | |
| | 6-Paramedicine Information | ✓ | | | | | | |
| | 7-3D picture | ✓ | | | | | | |
| | 8-2D picture | | | ✓ | | | | |
| | 9-Patient access | | ✓ | | | | | |
| User Interface | 1-User-friendly | | | | ✓ | | | |
| | 2-Easy to use | | | | ✓ | | | |
| | 3-Well-designed UI | | | ✓ | | | | |
| Public Health | 1-General training | | | ✓ | | | | |
| | 2-Patient's usage | | ✓ | | | | | |
| | 3-Helping assistant | | ✓ | | | | | |
| Workflow | 1-Online reservation | ✓ | | | | | | |
| | 2-Beneficial for dentist's workflow | | | ✓ | | | | |
| | 3-Beneficial for assistant's workflow | | | ✓ | | | | |
| | 4-Beneficial for patient's workflow | | | ✓ | | | | |

**Table 3.** The Result of the Interviews in 4 categories

| | | Beneficent | Ease of Use | UI | Interaction | Reliability | Satisfaction | Voluntary | IT knowledge |
|---|---|---|---|---|---|---|---|---|---|
| Beneficent | Pearson Correlation | 1 | .206 | .590** | .487* | .543* | .634** | .674** | .624** |
| | Sig. (2-tailed) | | .398 | .008 | .034 | .016 | .004 | .002 | .004 |
| | N | 19 | 19 | 19 | 19 | 19 | 19 | 19 | 19 |
| Ease of Use | Pearson Correlation | .206 | 1 | -.174 | -.088 | -.028 | -.239 | .079 | -.291 |
| | Sig. (2-tailed) | .398 | | .282 | .720 | .910 | .138 | .749 | .068 |
| | N | 19 | 40 | 40 | 19 | 19 | 40 | 19 | 40 |
| UI | Pearson Correlation | .590** | -.174 | 1 | .666** | .178 | .642** | .358 | .642** |
| | Sig. (2-tailed) | .008 | .282 | | .002 | .465 | .000 | .132 | .000 |
| | N | 19 | 40 | 40 | 19 | 19 | 40 | 19 | 40 |
| Interaction | Pearson Correlation | .487* | -.088 | .666** | 1 | .524* | .884** | .523* | .357 |
| | Sig. (2-tailed) | .034 | .720 | .002 | | .021 | .000 | .022 | .134 |

|  |  | Beneficent | Ease of Use | UI | Interaction | Reliability | Satisfaction | Voluntary | IT knowledge |
|---|---|---|---|---|---|---|---|---|---|
|  | N | 19 | 19 | 19 | 19 | 19 | 19 | 19 | 19 |
| Reliability | Pearson Correlation | .543* | -.028 | .178 | .524* | 1 | .643** | .664** | .154 |
|  | Sig. (2-tailed) | .016 | .910 | .465 | .021 |  | .003 | .002 | .528 |
|  | N | 19 | 19 | 19 | 19 | 19 | 19 | 19 | 19 |
| Satisfaction | Pearson Correlation | .634** | -.239 | .642** | .884** | .643** | 1 | .752** | .426** |
|  | Sig. (2-tailed) | .004 | .138 | .000 | .000 | .003 |  | .000 | .006 |
|  | N | 19 | 40 | 40 | 19 | 19 | 40 | 19 | 40 |
| Voluntary | Pearson Correlation | .674** | .079 | .358 | .523* | .664** | .752** | 1 | .298 |
|  | Sig. (2-tailed) | .002 | .749 | .132 | .022 | .002 | .000 |  | .215 |
|  | N | 19 | 19 | 19 | 19 | 19 | 19 | 19 | 19 |
| IT knowledge | Pearson Correlation | .624** | -.291 | .642** | .357 | .154 | .426** | .298 | 1 |
|  | Sig. (2-tailed) | .004 | .068 | .000 | .134 | .528 | .006 | .215 |  |
|  | N | 19 | 40 | 40 | 19 | 19 | 40 | 19 | 40 |

**. Correlation is significant at the 0.01 level (2-tailed).
*. Correlation is significant at the 0.05 level (2-tailed).

**Table 4.** variables correlations

| Dependent variable/ Independent variables | Unstandardized Coefficients | | Standardized Coefficients | t | Sig. | $R^2$ |
|---|---|---|---|---|---|---|
|  | B | Std. Error | Beta |  |  |  |

| | | | | | | | |
|---|---|---|---|---|---|---|---|
| Beneficent | UI | .334 | .145 | .340 | 2.297 | .035 | 0.981 |
| | Voluntary | .599 | .135 | .657 | 4.443 | .000 | |
| Ease of Use | Voluntary | 1.025 | .038 | .988 | 27.152 | .000 | 0.975 |
| UI | Interaction | .420 | .152 | .396 | 2.763 | .013 | 0.972 |
| | Beneficent | .611 | .146 | .600 | 4.192 | .001 | |
| Interaction | Satisfaction | 1.711 | .234 | 1.852 | 7.317 | .000 | 0.982 |
| | Voluntary | -.762 | .221 | -.873 | -3.448 | .003 | |
| Reliability | Voluntary | .815 | .037 | .982 | 21.751 | .000 | 0.961 |
| Satisfaction | Voluntary | .564 | .053 | .597 | 10.655 | .000 | 0.996 |
| | Interaction | .444 | .061 | .410 | 7.317 | .000 | |
| Voluntary | Ease of Use | .253 | .117 | .262 | 2.169 | .045 | 0.992 |
| | Interaction | -.342 | .168 | -.299 | -2.032 | .049 | |
| | Satisfaction | 1.090 | .246 | 1.031 | 4.433 | .000 | |
| IT knowledge | Beneficent | .437 | .175 | .501 | 2.501 | .023 | 0.972 |
| | UI | .422 | .172 | .491 | 2.456 | .025 | |

**Table 5.** Multivariate linear regression

**Figures:**

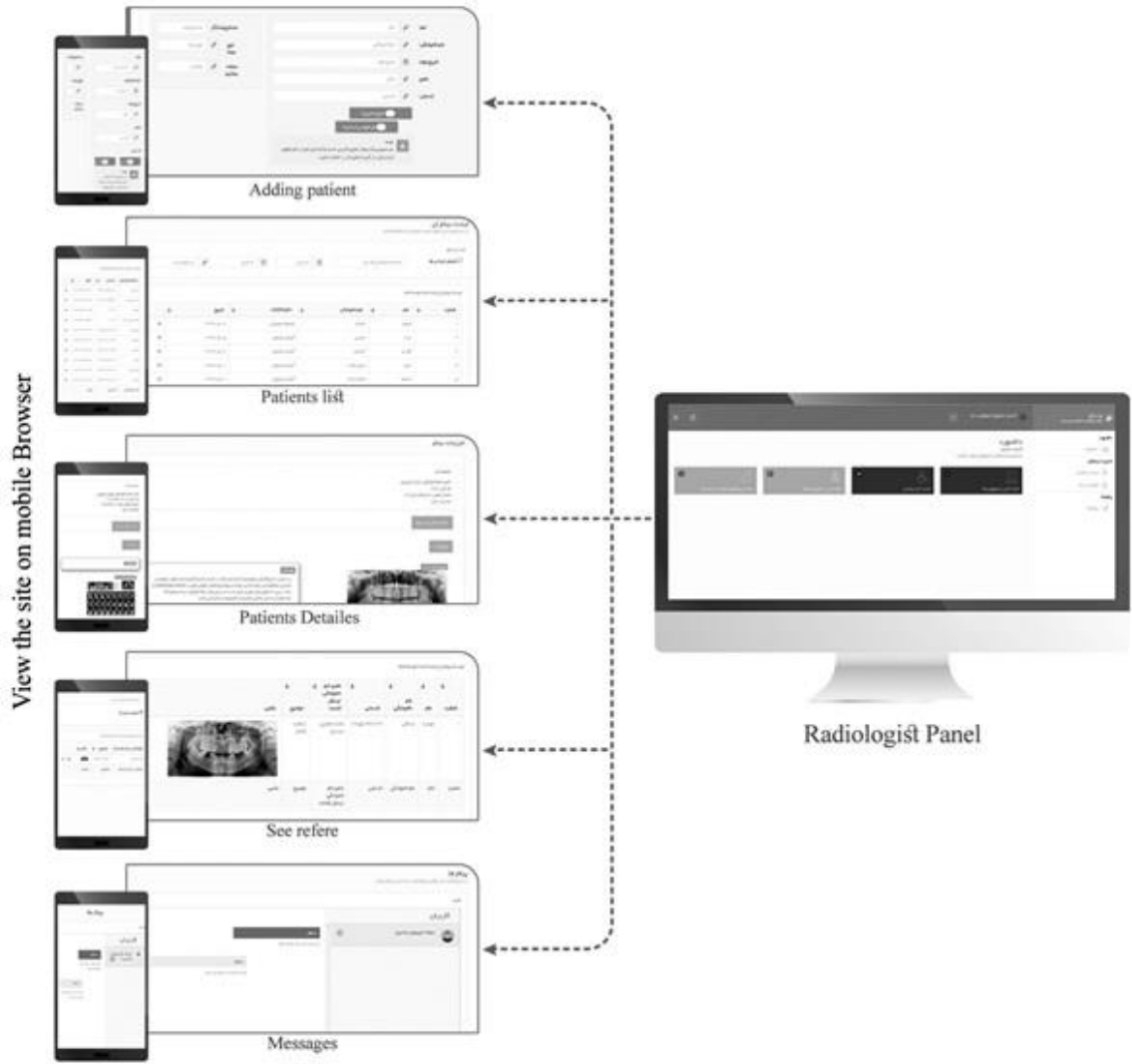

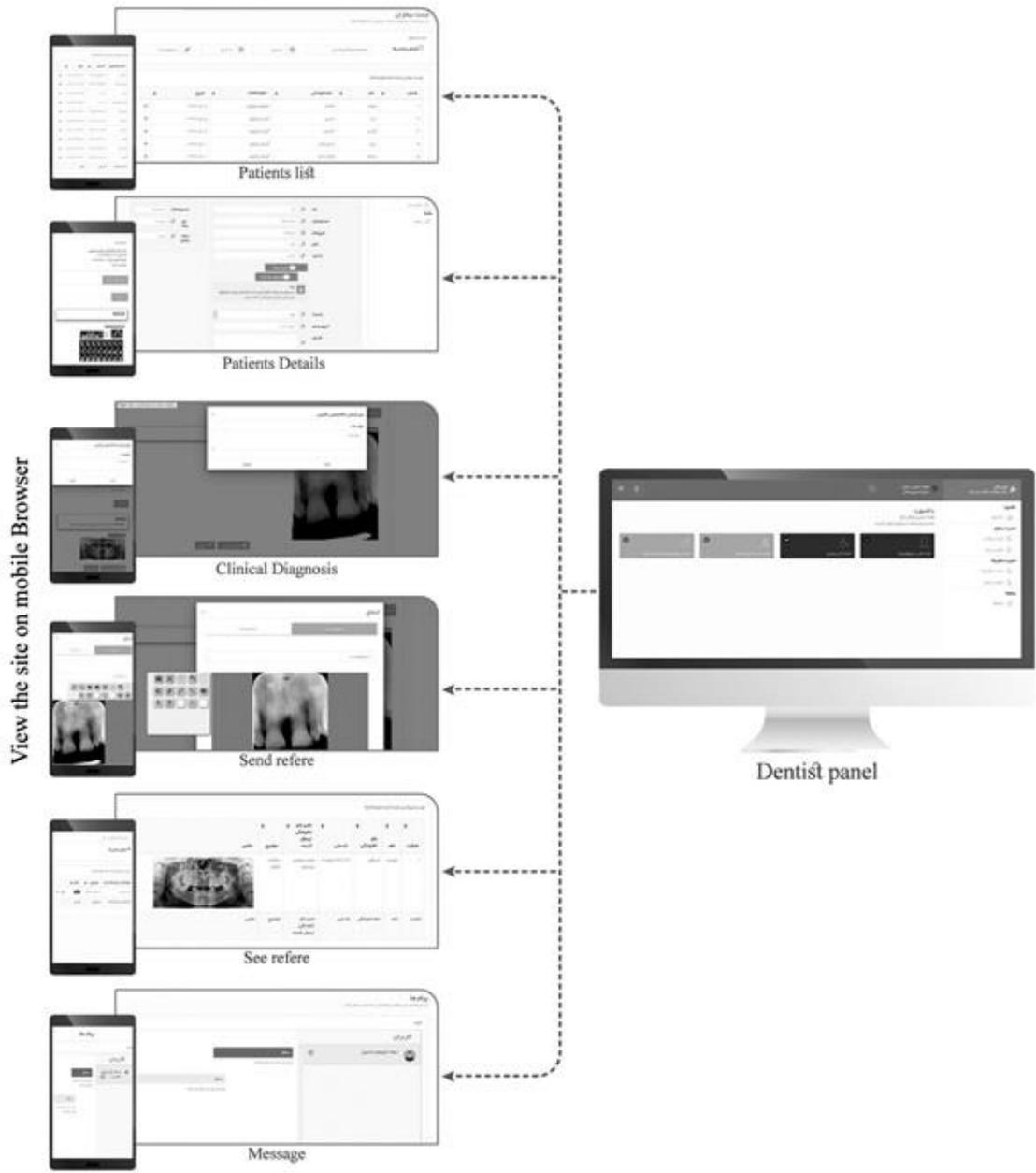

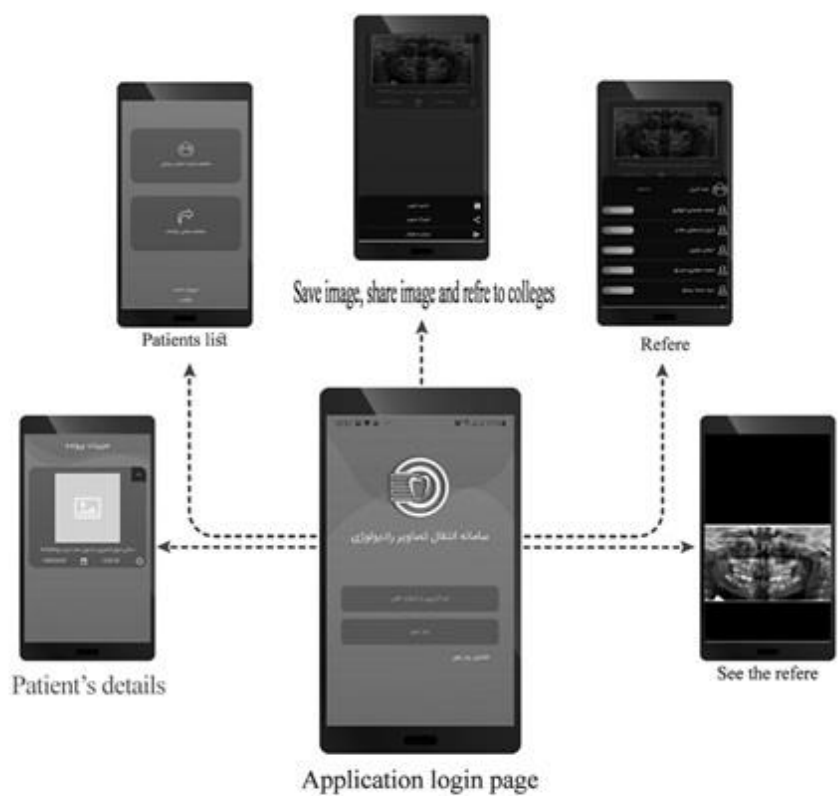

Figures captions:

1. Radiologist's dashboard
2. Dentist's dashboard
3. Android application